\documentclass[a4paper,fleqn]{cas-dc}

\usepackage[sort&compress,numbers]{natbib}
\usepackage[utf8]{inputenc}
\usepackage{textcomp}
\DeclareUnicodeCharacter{22EF}{\dots}
\DeclareUnicodeCharacter{0302}{\^}
\usepackage[version=4]{mhchem} %% Required for chemical equations
\usepackage{siunitx} %% Required to provide physical quantities as numbers and units together

\usepackage{soul}

\def\tsc#1{\csdef{#1}{\textsc{\lowercase{#1}}\xspace}}
\tsc{WGM}
\tsc{QE}
\tsc{EP}
\tsc{PMS}
\tsc{BEC}
\tsc{DE}

\begin{document}
\let\WriteBookmarks\relax
\def\floatpagepagefraction{1}
\def\textpagefraction{.001}
\shorttitle{Propylenidene}
\shortauthors{Laranjeira et~al.}

\title [mode = title]{Propylenidene: A Novel Metallic Carbon Monolayer with Unconventional Ring Topology}

\author[1]{José A. S. Laranjeira}
\affiliation[1]{
organization={Modeling and Molecular Simulation Group},
addressline={São Paulo State University (UNESP), School of Sciences}, 
city={Bauru},
postcode={17033-360}, 
state={SP},
country={Brazil}}
\cormark[1]
\cortext[cor1]{Corresponding author}
\credit{Conceptualization of this study, Methodology, Review and editing, Investigation, 
Formal analysis, Writing -- review \& editing, Writing -- original draft}
\author[2,3]{K. A. L. Lima}
\affiliation[2]{
organization={Institute of Physics},
addressline={University of Brasília}, 
city={Brasília },
postcode={70910‑900}, 
state={DF},
country={Brazil}}
\affiliation[3]{
organization={Computational Materials Laboratory, LCCMat, Institute of Physics},
addressline={University of Brasília}, 
city={Brasília},
postcode={70910‑900}, 
state={DF},
country={Brazil}}
\credit{Conceptualization of this study, Methodology, Review and editing, Investigation, 
Formal analysis, Writing -- review \& editing, Writing -- original draft}
\author[1]{Nicolas F. Martins}
\credit{Conceptualization of this study, Methodology, Review and editing, Investigation, 
Formal analysis, Writing -- review \& editing, Writing -- original draft}
\author[4]{{Luis A. Cabral}}
\credit{Investigation, Formal analysis, Resources, Writing -- review \& editing}
\affiliation[1]{organization={Department of Physics and Meteorology},
addressline={São Paulo State University (UNESP), School of Sciences}, 
city={Bauru},
% Uncomment if no comma needed between city and postcode
postcode={17033-360}, 
state={SP},
country={Brazil}}
\author[2,3]{L.A. Ribeiro Junior}
\credit{Conceptualization of this study, Methodology, Review and editing, Investigation, 
Formal analysis, Writing -- review \& editing, Writing -- original draft}
\author[1]{Julio R. Sambrano}
\credit{Conceptualization of this study, Methodology, Review and editing, Investigation, 
Formal analysis, Writing -- review \& editing, Writing -- original draft}

\begin{abstract}
Two-dimensional (2D) carbon allotropes have drawn significant interest owing to their impressive physical 
and chemical characteristics. Following graphene's isolation, a wide range of 2D carbon materials has 
been suggested, each with distinct electronic, mechanical, and optical traits. Rational design and 
synthesis of new 2D carbon structures hinge on experimentally reported precursors. Here, we present a 
2D carbon allotrope, propylenidene (PPD), originating from bicyclopropylidene. PPD forms a rectangular lattice with 3, 8, and 10-membered carbon rings. Density functional 
theory (DFT) simulations investigate its structural, electronic, mechanical, and optical properties. Our 
study shows PPD to be metallic. PPD exhibits 
absorption in the infrared and visible range, showing directional dependence in its response. Mechanically, 
PPD exhibits marked anisotropy; Young's modulus ($Y$) varies between 205.83 N/m and 164.46 N/m. These findings underscore 
the potential of this novel monolayer in applications such as energy storage, gas sensing, and optoelectronics.
\end{abstract}

% \includegraphics{figs/grabs.pdf}

%\item $\beta$-naphthyne is a novel porous 2D carbon allotrope.
%\item It features metallic-like behavior, ideal for energy storage with tunable charge transport.
%\item The structure consists of naphthyl units connected by octagonal rings.
%\item High mechanical anisotropy is observed in elasticity calculations.
%\item Strong optical absorption in IR and UV suggests optoelectronic potential.


\maketitle

\section{Introduction}

Two-dimensional (2D) carbon allotropes have garnered considerable attention due to their remarkable physical and chemical properties~\cite{Sharma2014pentahexoctite, Ram2018Tetrahexcarbon, Lu2013Two-dimensional, MORTAZAVI201857, SHAHROKHI2017205, MORTAZAVI2017344}. Since the isolation of graphene~\cite{Geim2009Graphene:}, a wide variety of 2D carbon-based structures have been proposed, each exhibiting distinct electronic, mechanical, and optical behaviors~\cite{Mortazavi2022A, Mortazavi2020Nanoporous, Li2017Graphene, laranjeira2024graphenyldiene, lima2024structural, BAFEKRY2020220, MORTAZAVI202040, SHAHROKHI201735, C9TC03513C}. Many of these allotropes feature porous architectures and have been predicted to perform well in applications such as gas sensing~\cite{martins2024ag, xie2023comparative}, metal-ion batteries~\cite{martins2024irida, zhao2025qpho}, and hydrogen storage~\cite{cheng2024metal, adithya2024decorated}.

Metallic 2D carbon allotropes are particularly attractive for energy-related applications due to their high electrical conductivity and rich $\pi$-electron systems. These materials can act as effective $\pi$-acceptors when decorated with alkali, transition, or post-transition metals, enabling strong charge transfer from the adsorbed species to the carbon framework. This interaction induces partial positive charges on the adsorbed atoms, stabilizing them on the surface and enhancing their reactivity.

Recent studies have demonstrated that metallic and conductive carbon allotropes, when decorated with metal atoms, can serve as efficient platforms for hydrogen storage, catalysis, and battery technologies. Darvishnejad \textit{et al.}~\cite{DARVISHNEJAD202440} investigated the decoration of a sp$^2$-hybridized 2D carbon framework known as PBCF-graphene with several transition metals (Sc, Ti, V, Cr, Mn), revealing that Cr-decorated structures could reversibly adsorb up to 17 H$_2$ molecules, achieving a capacity of 9.10 wt\% with adsorption energies in the optimal physisorption range. Similarly, Dewangan \textit{et al.}~\cite{DEWANGAN202337908} studied Li-decorated $\Psi$-graphene, a carbon allotrope containing pentagons, hexagons, and heptagons. Their results showed that a single Li atom could adsorb seven hydrogen molecules via polarization mechanisms, with an average adsorption energy of -0.31 eV/H$_2$, leading to a gravimetric capacity of 15.15 wt\%. 

Mahamiya et al.~\cite{MAHAMIYA202337898} extended this approach to PAI-graphene, composed of polymerized as-indacene units. They demonstrated that Li atoms could be stably anchored and each atom could bind up to four hydrogen molecules, resulting in a hydrogen capacity of 15.7 wt\%, surpassing the DOE targets. Bi et al.~\cite{BI202232552} focused on penta-octa-graphene (POG) and showed that decoration with Li or Ti yields stable systems capable of storing up to 9.9 wt\% of hydrogen, with adsorption energies between 0.14 and 0.95 eV. Cheng et al.~\cite{CHENG2024132405} explored M-graphene decorated with Li, Ca, and Sc, and found that single metal atoms could reversibly adsorb up to six H$_2$ molecules, with favorable binding energies and high gravimetric densities. They also showed that external electric fields could modulate the adsorption behavior.

In another work, Cheng et al.~\cite{CHENG202334164} investigated HOT-graphene, a Dirac semimetal with hexagonal, octagonal, and tetragonal rings. The authors showed that decoration with alkali and alkaline earth metals (Li, Na, K, Ca) led to ultrahigh hydrogen uptake capacities (up to 14.59 wt\%), and AIMD simulations confirmed the thermal stability of the metallized systems. Umadevi et al.~\cite{UMADEVI20241092} examined TPH-graphene nanoribbons and showed that Li and Na decoration significantly enhanced H$_2$ adsorption, yielding gravimetric capacities of 7.75 wt\% and 6.90 wt\%, respectively, along with stable charge transfer and polarization mechanisms.

In addition to hydrogen storage, conductive 2D carbon frameworks have proven valuable for battery applications. Martins et al.~\cite{MARTINS2025100910} employed OCD-graphene, a distorted carbon lattice with octagonal rings, as an anode material for sodium-ion batteries. Their simulations showed a high theoretical capacity (1339 mAh/g), low diffusion barrier (0.12 eV), and thermal stability confirmed by AIMD. Cai et al.~\cite{https://doi.org/10.1002/eem2.12127} studied Net-C18, a graphene-like metallic carbon sheet containing 5-, 6-, and 8-membered rings, and demonstrated that it could serve as a high-performance anode for lithium-ion batteries with a specific capacity of 403 mAh/g. Finally, Lima et al.~\cite{LIMA2025117235} introduced Petal-Graphyne, a metallic allotrope composed of 4-, 8-, 10-, and 16-membered rings, which showed excellent performance for both Li and Na-ion storage, with low migration barriers and theoretical capacities exceeding 1000 mAh/g.

On the other hand, a promising approach for designing new 2D carbon materials involves starting from experimentally known molecular precursors~\cite{Lu2013Two-dimensional, Pérez-Elvira2024Coronene‐Based, Zhao2019C-57}. By assembling these building blocks into extended networks, it becomes possible to tune hybridization, bonding geometry, and electronic structure rationally. This strategy enables the design of low-symmetry carbon lattices with tailored band gaps, mechanical anisotropy, and surface reactivity~\cite{Wan2024Design, Zhang2015Penta-graphene, Wei2017A}, facilitating their integration into nanoelectronics, energy storage, and catalysis platforms~\cite{Peng2018Carbon‐Supported, Hu2019Design, Shao2025Recent}.

Among such precursors, \textit{bicyclopropylidene} stands out as a highly strained hydrocarbon consisting of two fused cyclopropyl rings connected by a central double bond~\cite{de2000bicyclopropylidene}. Its unique geometry introduces significant ring strain and perturbs the conventional $\pi$-electron delocalization typically found in planar $sp^2$-hybridized systems~\cite{foerstner1998first, de1988bicyclopropylidene}. This reactivity has been explored in various transformations including rearrangements, cycloadditions, and polymerization~\cite{de2001spirocyclopropanated, wang2013photoinduced}, making it a compelling unit for constructing novel carbon frameworks.

In this work, we propose a new two-dimensional carbon allotrope named \textit{Propylenidene} (PPD), formed by organizing bicyclopropylidene units into a rectangular lattice comprising 3-, 8-, and 10-membered rings. Using density functional theory (DFT), we systematically investigate its structural, electronic, optical, and mechanical properties. The dynamical and thermal stability of PPD is confirmed through phonon dispersion, cohesive energy analysis, Born–Huang mechanical criteria, and \textit{ab initio} molecular dynamics (AIMD). The results suggest that PPD is a thermally robust and conductive 2D material with potential relevance for applications in hydrogen storage, alkali-ion batteries, and heterogeneous catalysis. While the present work focuses on the pristine monolayer, future studies will explore its performance as a tunable platform for energy conversion and storage applications.

\section{Methodology}

To investigate the structural, electronic, mechanical, and optical properties of the proposed Propylenidene (PPD) monolayer, we employed first-principles calculations based on density functional theory (DFT)~\cite{hohenberg1964inhomogeneous}, as implemented in the Vienna Ab Initio Simulation Package (VASP)~\cite{kresse1996efficient,kresse1999ultrasoft}. The exchange–correlation energy was treated using the generalized gradient approximation (GGA) with the Perdew–Burke–Ernzerhof (PBE) functional~\cite{perdew1996generalized}, and the core–valence electron interactions were described by the projector-augmented wave (PAW) method~\cite{blochl1994projector,kresse1999ultrasoft}.

A plane-wave kinetic energy cutoff of 520~eV was employed to ensure accurate total energy convergence. Structural optimization was performed using the conjugate-gradient algorithm until the Hellmann–Feynman forces on each atom were below 0.01~eV/\r{A}, and the total energy variation between consecutive ionic steps was less than 10$^{-5}$~eV. A vacuum region of 15~\r{A} was included along the out-of-plane direction to prevent interactions between periodic images. Brillouin zone sampling was performed using a $\Gamma$-centered Monkhorst–Pack grid of $12\times18\times1$ for static and electronic calculations, and a coarser $6\times9\times1$ mesh during structural relaxations.

To account for long-range van der Waals interactions, especially relevant in porous or open frameworks, Grimme’s DFT-D4 correction scheme~\cite{10.1063/1.4993215} was applied.

The thermodynamic stability of the PPD monolayer was assessed by computing the cohesive energy ($E_\text{coh}$), defined as:

\begin{equation}
E_{\text{coh}} = \frac{E_{\text{tot}} - \sum_i n_i E_i}{\sum_i n_i},
\end{equation}

where $E_{\text{tot}}$ is the total energy of the relaxed structure, $E_i$ is the total energy of an isolated atom of type $i$ (in this case, carbon), and $n_i$ is the number of atoms of that type in the unit cell.

The dynamical stability was confirmed through phonon dispersion calculations using density functional perturbation theory (DFPT), as implemented in the Phonopy code~\cite{togo2015first}. Thermal stability was further examined via \textit{ab initio} molecular dynamics (AIMD) simulations in the canonical (NVT) ensemble, using the Nosé–Hoover thermostat~\cite{nose1984unified,hoover1985canonical}, performed at multiple temperatures (300~K, 600~K, 900~K, and 1200~K) for a total simulation time of 5~ps and a time step of 1~fs. The scanning tunneling microscopy (STM) images were obtained using self-consistent calculations within the Quantum ESPRESSO package~\cite{Giannozzi_2009}, considering a bias voltage of $1.0$~V.

\section{Results and Discussion}

\begin{figure*}[pos=!htb]
    \centering
    \includegraphics[width=1\linewidth]{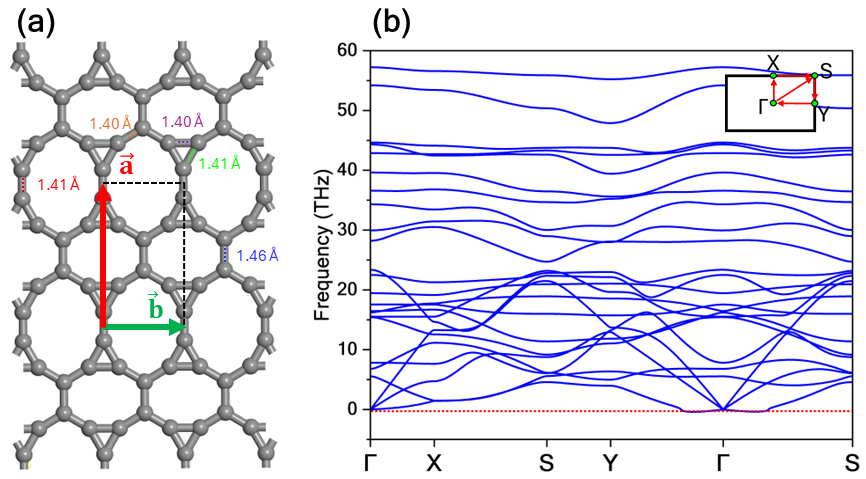}
    \caption{(a) Top view of atomic structure of propylenidene (PPD) monolayer. (b) Phonon dispersion 
    of PPD along high-symmetry paths of the Brillouin zone.}
    \label{fig:system}
\end{figure*}

PPD is represented by a rectangular unit cell that belongs to the $Pmmm$ (No. 47) space group, with lattice parameters $a = 6.70$~\r{A} and $b = 3.80$~\r{A} (see Fig.\ref{fig:system}(a)). This rectangular cell contains three distinct carbon atoms, namely C1 $(0.89746,~0.00000,~0.00000)$, C2 $(0.28464,~0.81572,~0.00000)$, and C3 $(0.60855,0.50000,0.00000)$. This atomic arrangement forms a 2D carbon framework comprising 3-, 8-, and 10-membered rings. The 10- and 8-membered pores have diameters of 5.24~\r{A} and 4.07~\r{A}, respectively.

The bond lengths in PPD are not uniform, reflecting the coexistence of different ring sizes and the angular strain of bicyclopropylidene motifs. Bonds within the cyclopropene-like rings range from 1.40~\r{A} to 1.41~\r{A}. Connections between the 3-membered rings and the larger 8-membered rings also exhibit lengths near 1.41~\r{A}, indicating partial double-bond character. In contrast, the bonds bridging the 10-membered rings extend up to 1.46~\r{A}.  Bond angles in the 
3-membered rings are near 60.0$^\circ$; in 8-membered rings, they range from  120.6$^\circ$ to 149.65$^\circ$; and in ten-membered rings, from 118.5$^\circ$ to 151.2$^\circ$.  The novel structure exhibits a cohesive energy ($E_\text{coh}$) of (-7.23 eV/atom), indicating the energy necessary to break down the solid into individual atoms. This value is comparable to that obtained for other carbon allotropes such as graphene (-7.68 eV/atom), T-graphene (-7.45 eV/atom), Graphenylene (-7.33 eV/atom), Graphenyldiene (-6.92 eV/atom), and Graphyne (-7.20 eV/atom), all calculated at the same theory level of this study.

The dynamical stability of PPD was evaluated by calculating the phonon band dispersion 
along the high-symmetry paths of the Brillouin zone, as shown in Fig. \ref{fig:system}(b). 
The absence of imaginary frequencies in the phonon spectrum confirms its stability. At the 
$\Gamma$-point, three acoustic phonon modes are observed. Most phonon branches are within the 
0–23 THz range, displaying multiple crossings, indicative of various thermal conductivity 
pathways. In the 27–44 THz range, the dispersion is reduced. As expected for two-dimensional materials, the phonon dispersion of PPD exhibits three acoustic branches: the longitudinal acoustic (LA), transverse acoustic (TA), and out-of-plane flexural acoustic (ZA) modes. The LA and TA modes show a linear dispersion near the $\Gamma$ point, reflecting in-plane vibrational motions typical of 2D systems. The third acoustic mode, the ZA mode, exhibits a parabolic dispersion near $\Gamma$, which is a well-known feature in 2D materials due to the restoring force being weaker for out-of-plane atomic displacements~\cite{YAN2023106731}. 

The highest vibrational frequency of PPD is approximately 57~THz ($\sim$1900~cm$^{-1}$), 
which is slightly higher than the LO/TO mode of graphene at the $\Gamma$-point 
($\sim$48~THz, $\sim$1580~cm$^{-1}$)~\cite{Falkovsky2007}. This increase is attributed to the presence of highly strained three-membered rings, which shift the optical modes to higher frequencies. The low-frequency acoustic modes (ZA, TA, LA) follow a similar trend to graphene but exhibit greater dispersion due to the lower lattice symmetry of PPD. In contrast to graphene, PPD displays a phonon gap between 44 and 50~THz, which originates from its unique 3--8--10 ring topology. The phonon dispersion was computed along the $\Gamma$--Y--S--X--$\Gamma$--S path in the rectangular Brillouin zone, shown in the inset, which differs from the hexagonal Brillouin zone characteristic of graphene.

\begin{figure*}[pos=!htb]
    \centering
    \includegraphics[width=1\linewidth]{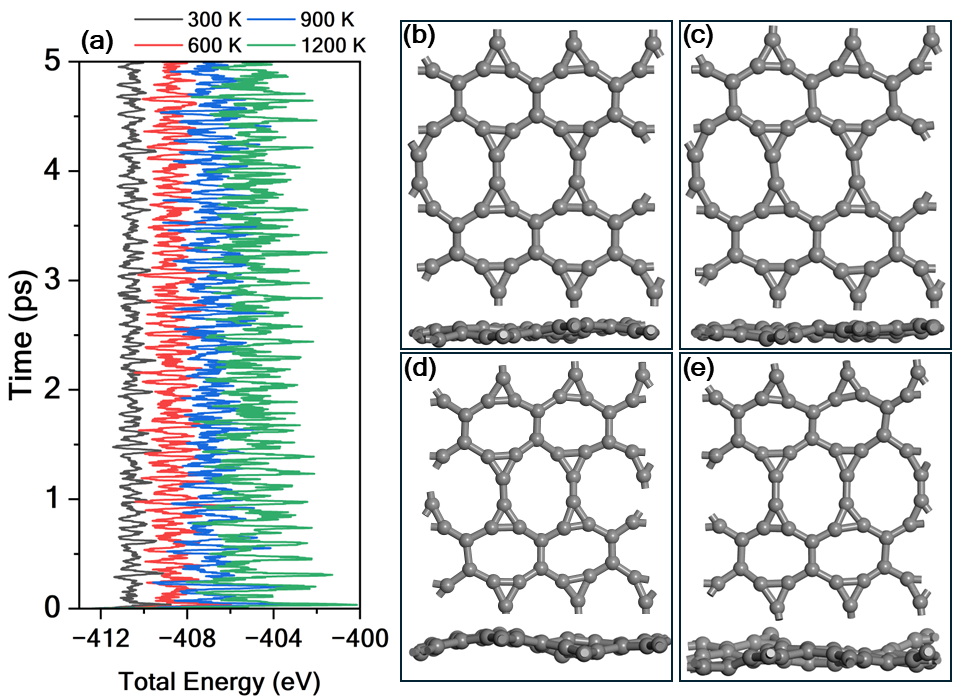}
    \caption{\textit{Ab initio} molecular dynamics (AIMD) simulations of PPD in the NVT ensemble over 5~ps with a 1~fs time step at four different temperatures. (a) Time evolution of the total energy at 300~K, 600~K, 900~K, and 1200~K, showing only small fluctuations and no signs of structural collapse. Final snapshots of the atomic configurations after the simulations at (b) 300~K, (c) 600~K, (d) 900~K, and (e) 1200~K.}
    \label{fig:md}
\end{figure*}

The thermal stability of PPD was evaluated through \textit{ab initio} molecular dynamics (AIMD) simulations in the canonical (NVT) ensemble for 5~ps with a 1~fs time step at four different temperatures: 300, 600, 900, and 1200~K. Fig.~\ref{fig:md}(a) shows the temporal evolution of the total energy, which exhibits only minor fluctuations at all temperatures, indicating the absence of structural collapse during the simulations. The corresponding final snapshots of the atomic configurations are displayed in Fig.~\ref{fig:md}(b–e). At 300~K [Fig.~\ref{fig:md}(b)], the monolayer retains its pristine geometry without noticeable distortions, with bond lengths fluctuating in the 1.35–1.50~\r{A} range. Increasing the temperature to 600~K [Fig.~\ref{fig:md}(c)] results in slight out-of-plane oscillations, with bond lengths varying between 1.36–1.53~\r{A}, but the overall topology and pore arrangement remain intact. At 900~K [Fig.~\ref{fig:md}(d)], moderate bond-angle deviations and enhanced rippling are observed, with bond lengths spanning 1.32–1.56~\r{A}, yet the framework preserves its connectivity. Even at 1200~K [Fig.~\ref{fig:md}(e)], despite more pronounced thermal distortions and increased undulation amplitudes, the bond length range (1.32–1.51~\r{A}) remains within values typical of stable $sp^2$-hybridized carbon networks. These results confirm that PPD maintains robust thermal stability up to at least 1200~K.

\begin{figure*}[pos=!htb]
    \centering
    \includegraphics[width=1\linewidth]{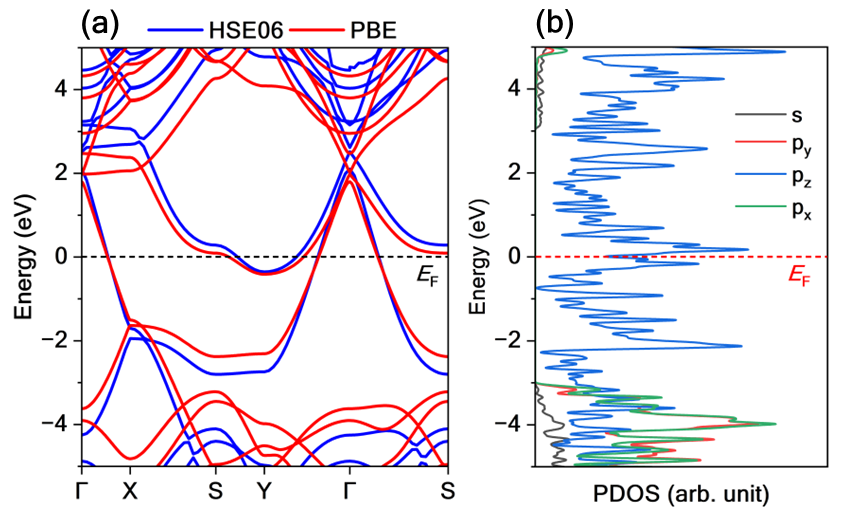}
    \caption{(a) Electronic band structure of propylenidene calculated using PBE (red curves) and HSE06 (blue curves) along the high-symmetry path $\Gamma$–$X$–$S$–$Y$–$\Gamma$–$S$. Both methods confirm metallic behavior, with several bands crossing the Fermi level ($E_F$) and an absence of a band gap. (b) Projected density of states (PDOS) showing the orbital contributions: $p_z$ orbitals dominate near $E_F$, consistent with $\pi$-electron delocalization responsible for the metallic conduction.}
    \label{fig:band}
\end{figure*}

Fig.~\ref{fig:band}(a) shows the electronic band structure of propylenidene calculated using the PBE (red curves) and HSE06 (blue curves) functionals. In both cases, the valence and conduction bands overlap at the Fermi level ($E_F$), confirming the metallic nature of the material with no discernible band gap. Several bands cross $E_F$ along different high-symmetry directions, indicating the presence of multiple conduction channels. Closer to the $\Gamma$ point, bands near $E_F$ exhibit steeper slopes, indicating lighter effective masses and potentially higher carrier velocities. While the general dispersion features are preserved across both methods, the HSE06 functional introduces a shift. It modifies the curvature of certain bands due to the improved treatment of exchange interactions. 

The projected density of states (PDOS), shown in Fig.~\ref{fig:band}(b), further elucidates the orbital contributions to the electronic structure. States in the vicinity of $E_F$ are dominated by $p_z$ orbitals, consistent with $\pi$-type delocalization across the planar carbon network, which facilitates metallic conduction. Contributions from $p_x$ and $p_y$ orbitals are smaller near $E_F$, while $s$-orbitals are mostly located at deeper energies, below $-4$~eV, indicating their minimal role in conduction processes. The dominance of $p_z$ character near the Fermi level suggests that the metallicity arises primarily from $\pi$-electron overlap, a feature common in conjugated carbon frameworks~\cite{LIMA2025417299, Laranjeira2025, LIMA2025117235, doi:10.1021/acsomega.5c04622}. 

\begin{figure}[pos=!htb]
    \centering
    \includegraphics[width=1\linewidth]{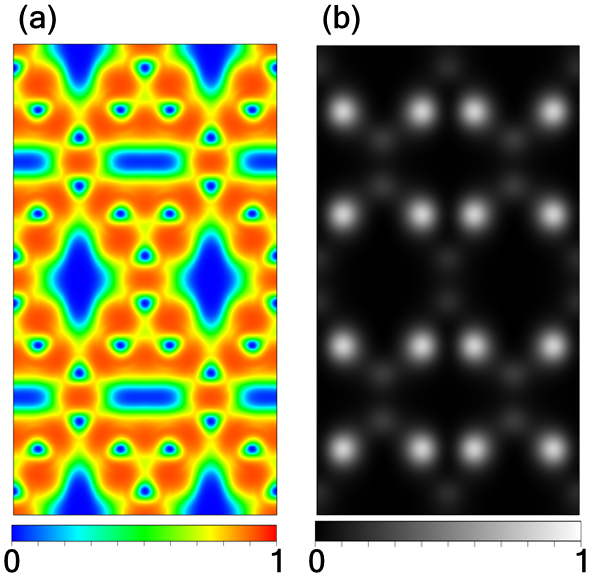}
    \caption{(a) Electron localization function (ELF) of PPD. (b) Simulated scanning tunneling 
    microscopy (STM) for PPD.}
    \label{fig:ELF}
\end{figure}

To better understand the charge distribution and bonding characteristics of PPD, we examined its electron 
localization function (ELF), depicted in Fig. \ref{fig:ELF}(a). The ELF effectively evaluates electron 
localization. A value of 1 indicates highly localized electrons, typical in covalent bonds or lone pairs. 
Conversely, a 0.5 reflects delocalized electrons akin to a homogeneous electron gas, and a value of 0 
represents areas with minimal electron density.

In PPD, the bicyclopropylidene motifs exhibit areas of high electron localization, as revealed by the electron localization function (ELF) map in Fig.\ref{fig:ELF}(a). This localization is primarily due to the angular strain imposed by the highly constrained 3-membered rings. In the bonds linking adjacent bicyclopropylidene units within the 8-membered rings, the ELF shows a noticeable reduction, suggesting a lower angular strain for these bonds. A similar trend is observed in the 10-membered rings, where the $\pi$ bonds of the bicyclopropylidene motifs display weaker localization compared to the cyclopropane-like rings themselves. 

Scanning tunneling microscopy (STM) simulations, shown in Fig.~\ref{fig:ELF}(b), provide further insight into the spatial distribution of the electronic states near the Fermi level. Bright spots in the simulated STM image correspond to regions with a high local density of states (LDOS), which are mainly associated with the cyclopropene-like rings within the PPD framework. These features could serve as experimental fingerprints for identifying the material.

\begin{figure}[pos=!htb]
    \centering
    \includegraphics[width=\linewidth]{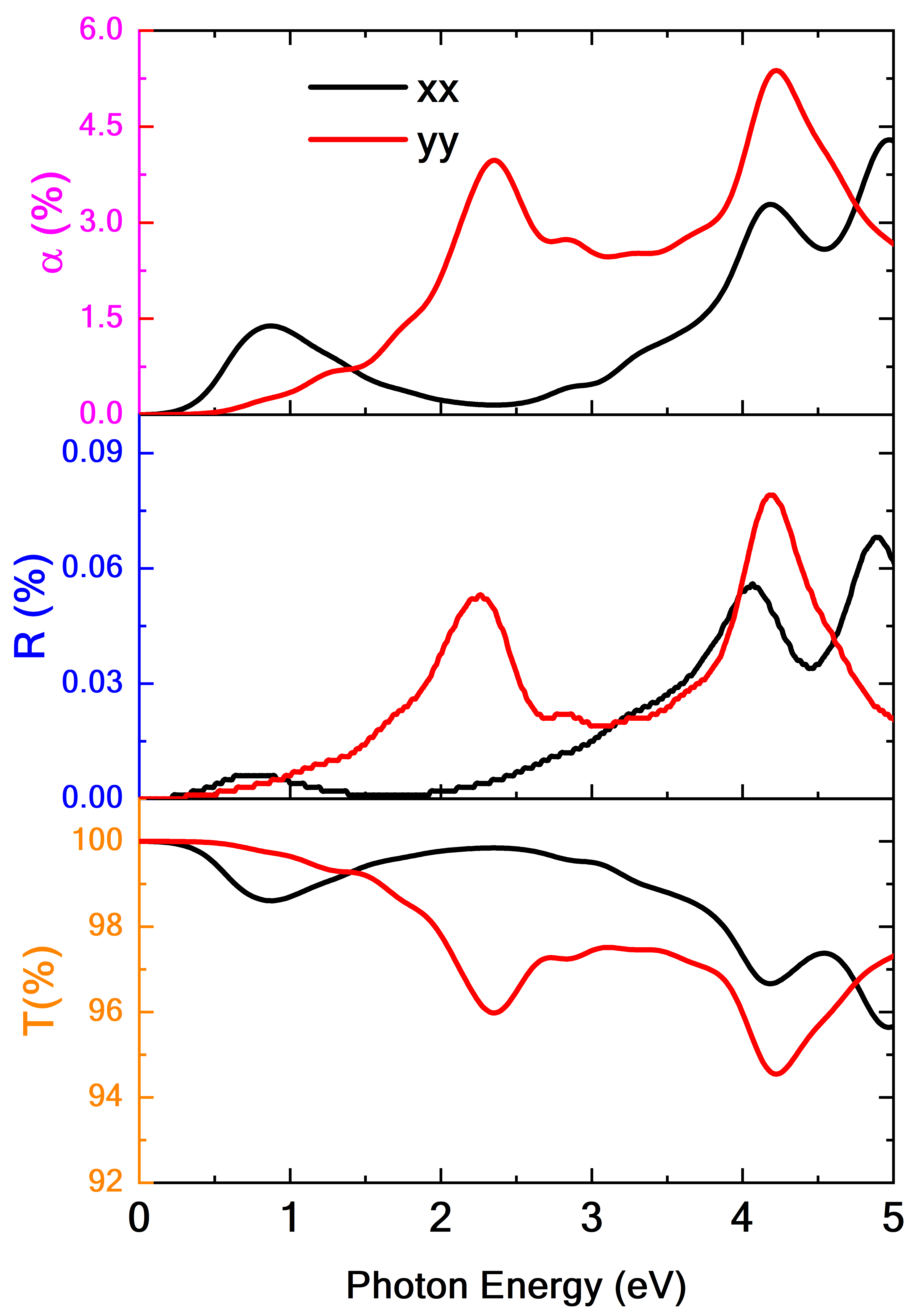}
    \caption{Optical properties of propylenidene for light polarized along the $x$ (black curves) and $y$ (red curves) crystallographic directions: (top) absorption coefficient $\alpha$, (middle) reflectivity $R$, and (bottom) transmittance $T$ as functions of photon energy.}
    \label{fig:optical}
\end{figure}

To investigate the optical properties of PPD, we calculated its absorption coefficient, as 
shown in Fig. \ref{fig:optical}. In the absorption spectra [Fig.~\ref{fig:optical}(top)], the first notable feature appears in the $xx$ polarization as a moderate peak of approximately 1.5\% located at $\sim$0.8~eV, within the infrared region. This low-energy response is likely associated with $\pi$–$\pi^*$ electronic transitions, consistent with the delocalized $p_z$ states near the Fermi level observed in the band structure and PDOS. The next pronounced peak emerges in the $yy$ polarization, reaching $\sim$4.0\% at $\sim$2.3~eV, within the visible range, indicating stronger light–matter coupling for this polarization and suggesting enhanced interband transitions along the $y$ crystallographic direction. At higher photon energies, both polarizations exhibit relatively strong absorption in the ultraviolet region, with maxima exceeding 4\%. 

The reflectivity spectra [Fig.~\ref{fig:optical}(middle)] remain low ($R < 0.07$) throughout the visible and near-UV regions, with peaks coinciding with the main absorption resonances, confirming that these features stem from direct interband transitions. The transmittance [Fig.~\ref{fig:optical}(down)] remains high ($T > 92\%$) across most of the studied range, with minima aligned with the absorption peaks. The clear difference in optical profile between the $xx$ and $yy$ components underscores the pronounced optical anisotropy of propylenidene, which could be exploited for polarization-sensitive optoelectronic applications.

The elastic constants of PPD were recalculated to refine the assessment of its mechanical properties, which are crucial for understanding structural stability and elastic response. The updated independent elastic constants are $C_{11} = 235.06$~N/m, $C_{22} = 225.76$~N/m, $C_{12} = C_{21} = 81.25$~N/m, and $C_{66} = 55.89$~N/m. According to the Born–Huang stability criteria for a rectangular lattice~\cite{born1940stability} — namely, $C_{11} > 0$, $C_{66} > 0$, and $C_{11}C_{22} > C_{12}^2$ — the PPD monolayer is mechanically stable.

\begin{figure*}[pos=!htb]
    \centering
    \includegraphics[width=1\linewidth]{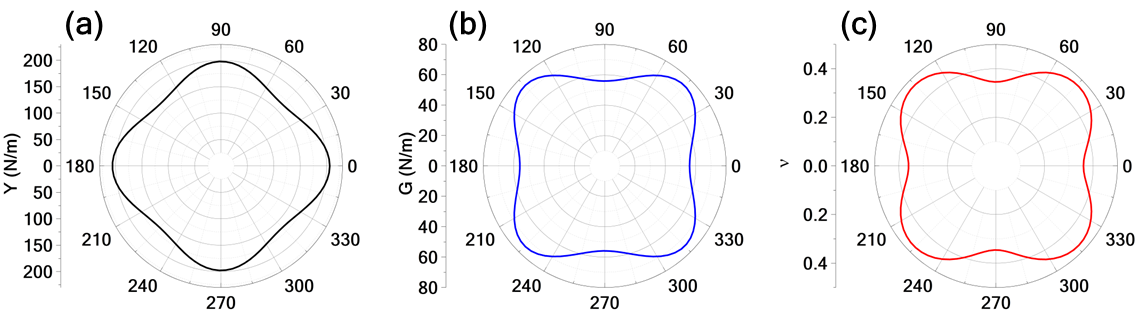}
    \caption{Polar plots of (a) Young’s modulus ($Y$), (b) Shear modulus ($G$), and (c) 
    Poisson's ratio ($\nu$)} for PPD.
    \label{fig:polar}
\end{figure*}

The polar plots of the in-plane mechanical properties of propylenidene (PPD)  [Fig.~\ref{fig:polar}(a–c)] reveal a distinct four-fold symmetry, consistent with the orthorhombic $Pmmm$ space group of the crystal lattice. This symmetry underscores the direct influence of the atomic-scale bonding topology on the macroscopic mechanical response, and reflects the intrinsic anisotropy of PPD. Young’s modulus ($Y$), which quantifies the resistance to uniaxial deformation, shows moderate anisotropy, varying between 164.46~N/m and 205.83~N/m 
(an anisotropy ratio of 1.25). The highest stiffness occurs along the $a$-axis and $b$-axis directions ($\theta = 0^\circ$ and $\theta = 90^\circ$), which are oriented parallel and perpendicular, respectively, to the bicyclopropylidene motifs. In contrast, the minimum stiffness ($Y_{\mathrm{min}} = 164.46$~N/m) is observed along the diagonal directions ($\theta = 45^\circ$, $135^\circ$, etc.), where applied stress induces torsional deformation of the C–C bonds that bridge two bicyclopropylidene units to form the octagonal rings. These bonds, measuring 1.46~\AA\ as previously highlighted, correspond to single bonds and thus possess greater rotational freedom, leading to a softer elastic response in these orientations.  

The shear modulus $G(\theta)$ [Fig.~\ref{fig:polar}(b)] and Poisson’s ratio $\nu(\theta)$ [Fig.~\ref{fig:polar}(c)] display the same four-fold rotational pattern, reflecting the fact that the orthorhombic symmetry imposes identical elastic responses under $90^\circ$ rotations. $G$ ranges from 55.89~N/m to 74.55~N/m, with an anisotropy ratio of 1.33. $\nu$ spans from 0.346 to 0.472, yielding an anisotropy ratio of 1.37.

The mechanical properties summarized in Table~\ref{tab:elastic_constants} highlight the balanced stiffness and anisotropy of the newly proposed propylenidene. Its Young’s modulus spans from 164.46~N/m to 205.83~N/m, exceeding that of $\alpha$-anthraphenylene while remaining below $\beta$- and $\gamma$-anthraphenylene, as well as more rigid allotropes such as graphene and penta-graphene. Compared to graphenylene, a porous carbon allotrope with experimentally reported synthesis~\cite{du2017new} and a Young’s modulus of 209.02~N/m (isotropic), propylenidene presents slightly lower stiffness but introduces mechanical anisotropy, which can be advantageous for direction-dependent applications. The maximum Poisson’s ratio (\(\nu_{\text{max}} = 0.47\)) is among the highest in the dataset, suggesting greater lateral strain accommodation under uniaxial loading compared to materials with low $\nu$, such as graphene (\(\nu = 0.17\)) or graphenylene (\(\nu = 0.27\)). In terms of shear modulus, propylenidene (\(G_{\text{max}} = 74.55\)~N/m, \(G_{\text{min}} = 55.89\)~N/m) surpasses $\alpha$-anthraphenylene and is comparable to $\beta$-anthraphenylene, indicating robust resistance to shear deformations with moderate anisotropy. Overall, the combination of relatively high stiffness, significant but controlled anisotropy, and elevated Poisson’s ratio positions propylenidene as a mechanically resilient and flexible 2D carbon allotrope, bridging the gap between highly rigid materials like graphene and more compliant, experimentally verified porous networks such as graphenylene.

\begin{table*}[pos=!htb]
\centering
\caption{Maximum and minimum values of Young's modulus (\(Y_{\text{max}}, Y_{\text{min}}\)) (N/m), Poisson's ratio (\(\nu_{\text{max}}, \nu_{\text{min}}\)), and shear modulus (\(G_{\text{max}}, G_{\text{min}}\)) (N/m) for several carbon monolayers, including the newly proposed propylenidene. The values were calculated employing the same theoretical framework employed to study the propylenidene.}
\begin{tabular}{lccc}
\hline
 & \(Y_{\text{max}}/Y_{\text{min}}\) & \(\nu_{\text{max}}/\nu_{\text{min}}\) & \(G_{\text{max}}/G_{\text{min}}\) \\
\hline
Propylenidene (this work) & 205.83/164.46 & 0.472/0.346 & 74.55/55.89 \\
Graphene~\cite{Geim2009Graphene:} & 345.42/345.42 & 0.17/0.17 & 147.60/147.60 \\
Penta-graphene~\cite{zhang2015penta}& 271.81/266.67 & -0.08/-0.10 & 151.21/144.98 \\
T-graphene~\cite{sheng2011t} & 293.90/148.02 & 0.58/0.16 & 148.02/126.57 \\ 
$\gamma$-anthraphenylene~\cite{LIMA2025417299} & 281.00/158.27 & 0.47/0.25 & 79.67/57.35 \\
$\beta$-anthraphenylene~\cite{LIMA2025417299} & 215.50/167.53 & 0.34/0.23 & 76.26/64.35 \\
$\alpha$-anthraphenylene~\cite{LIMA2025417299} & 169.94/127.89 & 0.40/0.26 & 65.29/52.87 \\ 
PHE-graphene\cite{zeng2019new}& 262.29/262.29 & 0.26/0.26 & 103.91/103.91 \\
Graphenylene~\cite{song2013graphenylene, du2017new} & 209.02/209.02 & 0.27/0.27 & 82.11/82.11 \\
Graphenyldiene~\cite{laranjeira2024graphenyldiene} & 122.47/122.47 & 0.35/0.35 & 45.29/45.29 \\
\hline
\end{tabular}
\label{tab:elastic_constants}
\end{table*}

\section{Conclusion}

This study presents a comprehensive analysis of the structural, electronic, mechanical, and optical 
properties of the newly proposed porous 2D carbon allotrope, propylenidene (PPD). Our first-principles 
calculations confirm its stability, supported by phonon dispersion and molecular dynamics simulations. 
The 2D structure, composed of 10-8-3 carbon-membered rings, introduces distinctive pores with 
diameters of 5.24 \r{A} and 4.07 \r{A}, respectively.

PPD demonstrates metallic behavior, with a large contribution of $\pi$ p$_z$ states closer to the Fermi level. 
Optically, PPD exhibits absorption in the infrared and visible range, showing directional 
dependence in its response. Mechanically, the material presents high anisotropy for Young's 
modulus, shear modulus, and  
Poisson's ratio, with significant variation across different crystallographic directions. 
Overall, PPD emerges as a promising material with a combination of stability, metallicity, and tailored mechanical and optical properties, making it an exciting candidate for 
future research and practical applications in advanced nanotechnology and materials science.

\section*{Data access statement}
Data supporting the results can be accessed by contacting the corresponding author.

\section*{Conflicts of interest}
The authors declare no conflict of interest.

\section*{Acknowledgements}
This work was supported by the Brazilian funding agencies Fundação de Amparo à Pesquisa do Estado 
de São Paulo - FAPESP (grant no. 2022/03959-6, 2022/14576-0, 2020/01144-0, 2024/05087-1, 
2024/19996-3 and 2022/16509-9), and National Council for Scientific, Technological Development 
- CNPq (grant no. 307213/2021–8). L.A.R.J. acknowledges the financial support from FAP-DF grants 
00193.00001808/2022-71 and $00193-00001857/2023-95$, FAPDF-PRONEM grant 00193.00001247/2021-20, 
PDPG-FAPDF-CAPES Centro-Oeste 00193-00000867/2024-94, and CNPq grants $350176/2022-1$ and 
$167745/2023-9$. The computational facilities were supported by resources supplied by the 
``Centro Nacional de Processamento de Alto Desempenho em São Paulo (CENAPAD-SP)'' and 
CENAPAD-RJ (SDumont).

\section*{Declaration of generative AI and AI-assisted technologies in the writing process}
During the preparation of this work the authors used Writefull in order to improve 
the readability and language of the manuscript. After using this tool/service, the authors 
reviewed and edited the content as needed and take full responsibility for the content 
of the published article.

\printcredits

\bibliography{cas-refs}

\end{document}